\documentclass{article}

\usepackage[preprint]{neurips_2021}

\usepackage[utf8]{inputenc} 
\usepackage[T1]{fontenc}    
\usepackage{hyperref}       
\usepackage{url}            
\usepackage{booktabs}       
\usepackage{amsfonts}       
\usepackage{nicefrac}       
\usepackage{microtype}      
\usepackage{xcolor}         

\usepackage{graphicx}
\usepackage{amsmath}
\usepackage{array}
\usepackage{sidecap}

\title{Learning to infer in recurrent biological networks}

\author{%
  Ari S. Benjamin \& Konrad P. Kording \\
  Department of Bioengineering\\
  University of Pennsylvania\\
  Philadelphia, PA 19104 \\
  \texttt{aarrii@seas.upenn.edu} \\
  }

\begin{document}

\maketitle

\begin{abstract}
A popular theory of perceptual processing holds that the brain learns both a generative model of the world and a paired recognition model using variational Bayesian inference. Most hypotheses of how the brain might learn these models assume that neurons in a population are conditionally independent given their common inputs. This simplification is likely not compatible with the type of local recurrence observed in the brain. Seeking an alternative that is compatible with complex inter-dependencies yet consistent with known biology, we argue here that the cortex may learn with an adversarial algorithm. Many observable symptoms of this approach would resemble known neural phenomena, including wake/sleep cycles and oscillations that vary in magnitude with surprise, and we describe how further predictions could be tested. We illustrate the idea on recurrent neural networks trained to model image and video datasets. This framework for learning brings variational inference closer to neuroscience and yields multiple testable hypotheses. 
\end{abstract}

\section{Introduction}

Animals can predict future sensations, are surprised when predictions are violated, and learn from such surprises to predict better. It is standard to formalize perceptual expectations as probability distributions over possible new sensations \citep{Fiser2010}, which are called an organism's internal models. Internal models carry clear benefits to a perceiving brain and are ubiquitous in computational neuroscience as theories of how the brain ought to perceive \citep{wolpert1995internal}.

When new observations are not predicted by the internal models, the brain must integrate them into representations at various levels. This requires interpreting the low-level sensations in the context of its generative model. In this Helmholtzian framework \citep{Helmholtz1925, knill1996perception}, perception is understood as inference over an internal model of the world. Many psychophysical and physiological experiments support this understanding, and as a result this idea is now central to the modern theory of perception \citep{poggio1987computational, mumford1994neuronal,hinton1997generative, friston2005theory, Yuille2006,Berkes2011a, kleinschmidt2015robust,dasgupta2020theory}.

Just as an internal model must be learned, it is often argued that the brain learns to infer via a recognition model instantiated in bottom-up synapses \citep{hinton1995wake,dasgupta2020theory}. Though multiple models of learning have been proposed \citep{hinton1995wake,friston2005theory,rezende2014bio}, it is still debated which algorithm for this task could most directly and unambiguously explain synaptic plasticity. We wish to identify additional algorithms in this family that are both consistent with experimental knowledge and that work well to train networks like those in the cortex, which at a minimum are hierarchical, recurrent, and nonlinear.

There now exist algorithms that work quite well when the hierarchical and nonlinear conditions are met \citep{child2020very}, but these are not compatible with recurrence within the recognition and generative models. Recurrence causes neurons within a layer to be mutually dependent. However, algorithms like the Wake-Sleep algorithm \citep{hinton1995wake} and variational autoencoders with the usual diagonal Gaussian conditional distributions \citep{kingma2013auto,rezende2014stochastic} assume that neurons in one layer are independent from one another, conditioned on the previous layer. Because a mean-squared objective implies Gaussian-distributed error, this assumption also underlies the hypothesis within theoretical neuroscience that each feedback synapse learns by predicting postsynaptic activity \citep{friston2005theory,rezende2014bio,urbanczik2014learning}. These algorithms cannot be easily generalized while remaining biologically plausible. Ultimately, the problem posed by recurrence is that synapses evaluating the likelihood of postsynaptic activity now also require nonlocal information about many other neurons' activity. If the brain performs Bayesian inference over neural activity, it must learn with algorithms compatible with dependencies between neighboring neurons.

The difficulty posed by inter-dependencies can be resolved if the brain contains circuits designed to help align more general probability distributions of activity. Here, we consider the possibility that local circuits align these distributions adversarially, i.e. with teaching signals that can be interpreted as discriminators. This hypothesis leverages recent demonstrations in the generative modeling community that adversarial approaches can be used to learn Bayesian inference over a generative model \citep{donahue2016adversarial,dumoulin2016adversarially, pu2017adversarial, srivastava2017veegan,donahue2019large}. 
In these papers, feedforward and feedback connections both learn by opposing a discriminator of whether pairs $(x,z)$ of inputs and hidden representations are driven top-down or bottom-up. If these two phases can be observed separately, learning a discriminator is a simple yet powerful way to learn to perform Bayesian inference over a generative model.

In this paper we propose an adversarial model of sensory learning in the cortex. The remainder of the paper is as follows. Section 2 reviews variational inference and the difficulty posed by recurrence. Section 3 describes our adversarial model. We describe how two timescales could be used for learning: one in which a long period of stimulus-evoked responses is followed by an equal period of self-generated activity, as in the Wake-Sleep algorithm \citep{hinton1995wake}, and another in which areas quickly alternate between a bottom-up drive and a fast top-down reconstruction or prediction of the immediate future. Section 4 lays out the biological predictions of our hypothesis. In Section 5 we implement the model on example tasks to illustrate the algorithm's compatibility with recurrent architectures.
   
\section{Background: variational inference and its biological implementation}

The purpose of variational inference is to learn mappings between inputs $\mathbf{x}$ and a learned representation $\mathbf{z}$ that are reciprocal in a specific Bayesian sense. In our model both  $\mathbf{x}$ and $\mathbf{z}$ are taken to be vectors of neural activities. The top-down network generates a distribution over $\mathbf{x}$ for each $\mathbf{z}$, which we can write as $p_\theta(\mathbf{x}|\mathbf{z})$. This depends on the parameters $\theta$. Together with a prior distribution over $\mathbf{z}$, this implies a \textit{joint} distribution $p_\theta(\mathbf{x},\mathbf{z})$. One objective of learning is that the network should model the observed inputs, in the sense that the marginal $p_\theta(\mathbf{x})$ equals the input distribution $q(\mathbf{x})$.

Conversely, the bottom-up network maps an observed $\mathbf{x}$ to a distribution over $\mathbf{z}$,  $q_\phi( \mathbf{z}|\mathbf{x})$, which depends on the learnable parameters $\phi$. Together with the true input distribution $q(x)$, this implies a joint distribution $q_\phi(\mathbf{x},\mathbf{z})$. A second objective of learning is that $q_\phi( \mathbf{z}|\mathbf{x})$ should infer the posterior distribution of which $\mathbf{z}$ could have generated an observed $\mathbf{x}$ under the internal model, i.e. $p_\theta(\mathbf{z}|\mathbf{x})$.

Both of these objectives are met when $q_\phi(\mathbf{x},\mathbf{z})=p_\theta(\mathbf{x},\mathbf{z})$, i.e. when the joint probability distribution of \textit{(real data, inferred representation)} pairs matches the joint distribution of \textit{(sampled representation, generated data)} pairs. Learning in this framework is thus a distribution-matching problem. 

Most variational inference algorithms minimize the Kullbeck-Leibler (KL) divergence between these two joint distributions. The KL divergence between the two joint distributions is:
\begin{equation} \label{eq:free-energy}
D_{KL}(q_\phi\|p_\theta) = \mathop{\mathbb{E}}_{q_\phi( \mathbf{x},\mathbf{z})}\bigg[\log \frac{
q_\phi( {\mathbf{x}}, \mathbf{z})
}{
p_\theta( \mathbf{x},\mathbf{z})
}\bigg].
\end{equation}
This expression is sometimes referred to as the variational free energy. Note that this is typically written using the conditional $q_\phi(\mathbf{z}|\mathbf{x})$, whereas we use the joint $q_\phi( \mathbf{x},\mathbf{z})$ to emphasize the distribution-matching interpretation. This does not change optimization because the value of the equation changes only by the entropy of the inputs $\mathbf{x}$, which is independent of learnable parameters. More general introductions to variational inference can be found in various textbooks \citep{dayan2001theoretical,bishop2006pattern}.

\subsection{Can the cortex minimize $D_{KL}$?}

Synaptic plasticity can reduce an objective in three ways: if it is a function of local information, if it computed elsewhere and transmitted as a reward signal, or if its derivatives are computed and propagated through a network. Since changing any synapse affects the activity of all downstream neurons, the purely local option is not available and $D_{KL}$ must be computed and communicated to distant synapses for learning.

\begin{SCfigure}\label{fig:whatswrong}
\includegraphics[width=.5\linewidth]{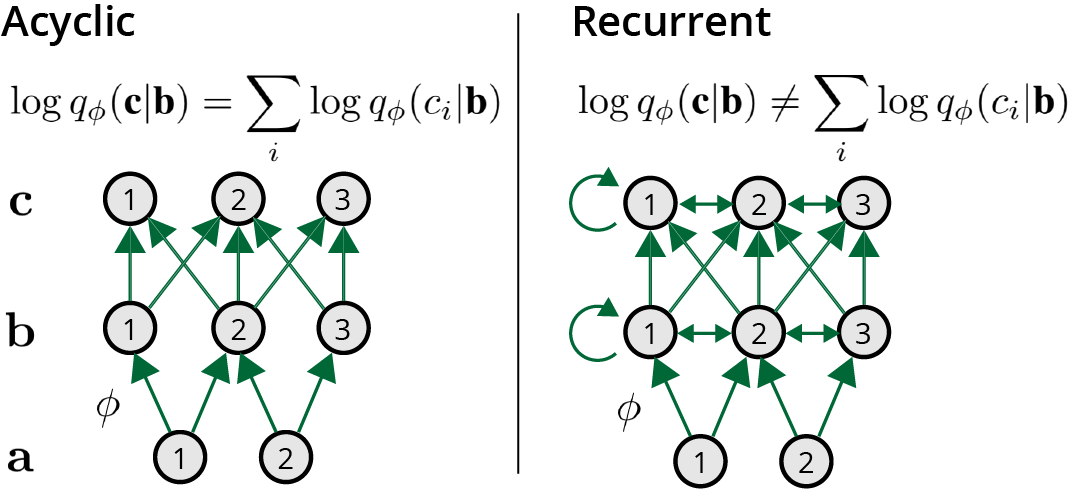} 
\caption{The problem of recurrence. \textit{Left: } When neural activity depends only on a previous layer, their conditional probability given that layer factorizes into single-neuron terms. In this case the objective $D_{KL}$ becomes a sum of single-neuron prediction errors and entropies. \textit{Right:} Local recurrence prevents conditional probabilities from factorizing. Thus $\log p_\phi(\mathbf{c|b})$ is not locally defined and depends on all activities in $\mathbf{c}$.}
\end{SCfigure}

Computing $D_{KL}$ is difficult in the general case because it is a quantity that depends on the activity of all neurons. When no assumptions about connectivity can be made, the probability of observing any one neuron's activity changes with knowledge of other neurons' activities.
This transport problem prohibits an exact calculation of $D_{KL}$ except in certain architectures.

\subsection{$D_{KL}$ is calculable in an acyclic network } \label{sec:acyclic}

Most algorithms for variational inference with a recognition model that have been proposed as biological hypotheses assume a hierarchy of layers with no local interconnectivity within each layer (Fig. 1).  This means neurons in any layer are independent of one another, conditioned on the activity of the previous layer. In this condition the bottom-up and top-down connections both form directed acyclic graphs, which means that $\log q_\phi( \mathbf{x},\mathbf{z})$ and $\log p_\theta( \mathbf{x},\mathbf{z})$ are sums of log-probabilities of single neural activities: 
\begin{equation} \label{eq:DAG-logjointq}
\log q_\phi( \mathbf{x},\mathbf{z})=
\sum_i \log q_\phi(z_i|\mathrm{pa}_i^q)
\hspace{20pt}
\bigg|
\hspace{20pt}
\log p_\theta( \mathbf{x},\mathbf{z})=
\sum_i \log p_\theta(z_i|\mathrm{pa}_i^p)
\end{equation}
Here $\mathrm{pa}_i^q$ represents the parents (neurons presynaptic to) neuron $i$ in the inference graph, and $\mathrm{pa}_i^p$ represents its parents in the generative graph.

With this assumption, $D_{KL}$ becomes a sum of single-neuron prediction errors and entropies. These can be computed locally at each synapse. 
Note that since neurons affect downstream activity, these prediction errors and entropies must be summed across all neurons and that sum (the KL divergence) or its derivatives broadcast for optimal learning (as in \citet{rezende2014bio}) or communicated via message-passing. This strategy of connectivity restrictions and factorization underlies the bulk of variational inference algorithms proposed as models of the brain.

\paragraph{Example: a multilayer perceptron.} For a neural network with probabilistic neurons, $N$ layers, and strictly feedforward or feedback passes, the joint distributions factor by layer and additionally by neurons.
In this case $D_{KL}$ can be written as such, with $z_{i,j}$ denoting the $j$th neuron in the $i$th layer:
\begin{equation}\label{eq:KL_MLP}
D_{KL}(q_\phi\|p_\theta)
=
\mathop{\mathbb{E}}_{  q_\phi(\mathbf{x}, \textbf{z})}
\bigg[
\sum_j\log \frac{p_\theta(z_{N,j})}{q_\phi(z_{N,j}|\textbf{z}_{i-1})}+
\sum_i\sum_j\log \frac{p_\theta(z_{i,j}|\textbf{z}_{i+1})}{q_\phi(z_{i,j}|\textbf{z}_{i-1})}+
\sum_j\log \frac{p_\theta(x_j|\textbf{z}_{i+1})}{q(\mathbf{x})}
\bigg]
\end{equation}
Thus for a forward-only MLP, the variational objective is the sum of many locally-computed terms.

\paragraph{The problem posed by recurrence:}
If recurrence participates in the recognition or generative models, the key conditional probability distributions do not factorize into a sum of single-neuron terms. Imagining a recurrent population with activity $\mathbf{z}_i$ that receives input from the population $\textbf{z}_{i-1}$, then in general $\log p(\textbf{z}_i|\textbf{z}_{i-1})\neq\sum_j\log p(z_{i,j}|\textbf{z}_{i-1})$. Any one synapse from neuron $z_{i-1,j}$ to neuron $z_{i,k}$ cannot simply learn to predict the postsynaptic neuron but must observe all neurons in $\textbf{z}_i$ to change optimally. Unfortunately the number of fast signals available to synapses for learning are much fewer than the number of neurons in a local population. 
The solution to this problem is to find signals available for plasticity that \textit{aggregate} local activity and indicate the dissimilarity of population activity from the self-generated joint distribution. 

\begin{SCfigure}\label{fig:algorithms}
\includegraphics[width=.5\linewidth]{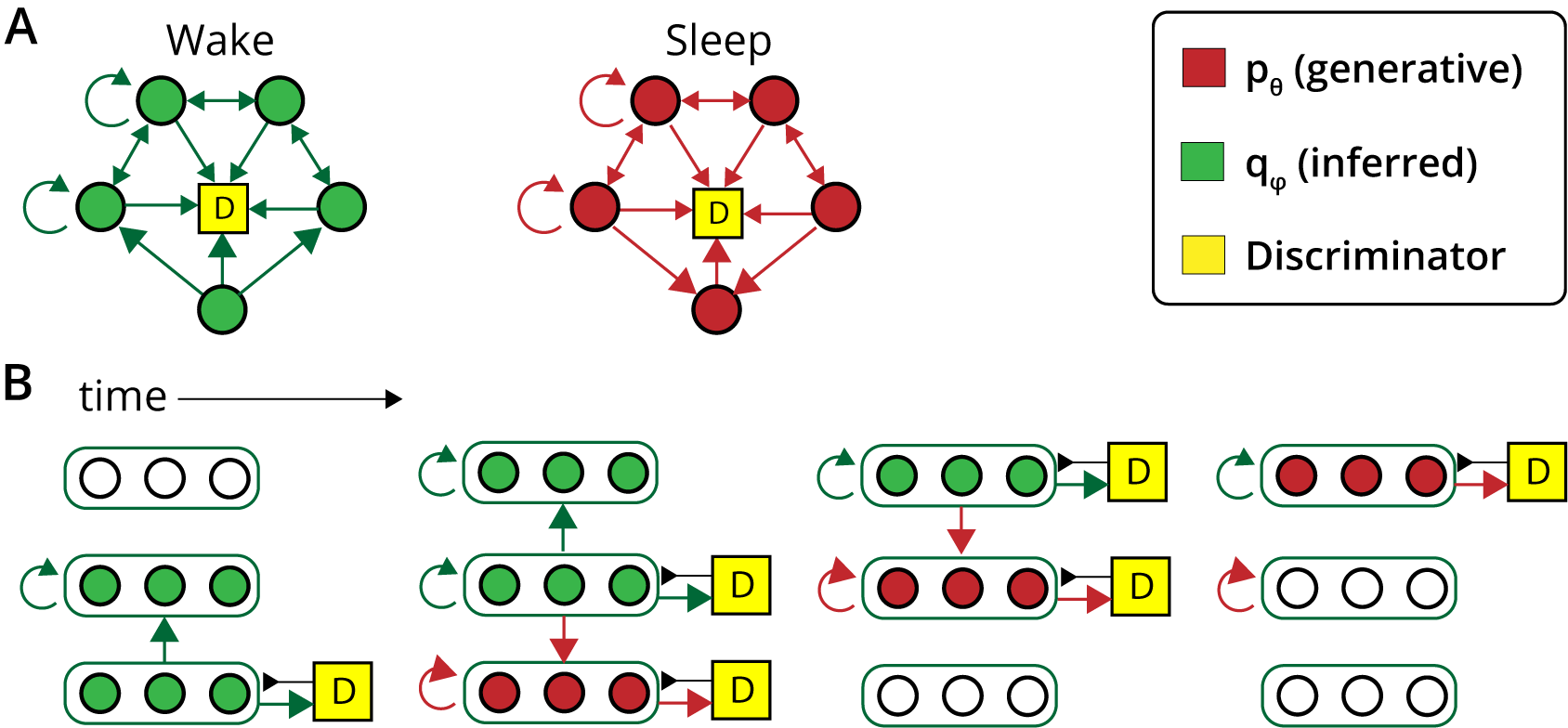} 
\caption{\textbf{A)} In an adversarial Wake/Sleep, neurons in a recurrent network are alternately driven by external inputs and spontaneous activity. A discriminator learns to classify the phase of activity and serves as a teaching signal to align them. \textbf{B)} A discriminator can also observe a wave of stimulus-driven activity ascending a hierarchy. Generative feedback is passed down one layer to drive activity, and a local discriminator observes both phases.}
\end{SCfigure}

\section{Two adversarial algorithms for sensory learning}

Here we consider two adversarial algorithms adapted from the generative modeling literature as hypotheses of cortical learning. Each is compatible with local recurrence. Though we introduce them for static inputs, these can easily be extended to data with a temporal dimension (see Section 5).

\subsection{Comparing wake and sleep distributions}
The proximate goal of variational Bayesian inference is to align the joint distributions of stimulus-driven activity and self-generated activity, $ q_\phi( \mathbf{x},\mathbf{z})$ and $p_\theta( \mathbf{x},\mathbf{z})$. Following the Wake-Sleep algorithm \citep{hinton1995wake}, we can nickname $ q_\phi( \mathbf{x},\mathbf{z})$ as the Wake distribution and $p_\theta( \mathbf{x},\mathbf{z})$ as the Sleep distribution. 

To adversarially align $q_\phi$ and $p_\theta$, one can introduce a discriminator $D(\mathbf{x},\mathbf{z})$ which is a function of the entire network. The discriminator is an estimate of whether the network state is more likely self-generated or stimulus-evoked. The discriminator learns by maximizing the objective:
\begin{equation}\label{WS}
\mathcal{L}_{WS}=
\mathop{\mathbb{E}}_{ q_\phi( \mathbf{x},\mathbf{z})}\bigg[D( {\mathbf{x}}, \mathbf{z})\bigg]
-\mathop{\mathbb{E}}_{ p_\phi( \mathbf{x},\mathbf{z})}\bigg[D( {\mathbf{x}}, \mathbf{z})\bigg]
\end{equation}
The particular formulation of Eq. \ref{WS}, the difference of the average of $D$ between phases, is but one among a large number of adversarial objectives suitable for matching distributions \citep{nowozin2016f}. It is that of the Wasserstein GAN, which can be derived from a minimization of the Wasserstein-1 distance between probability distributions \citep{arjovsky2017wasserstein} and requires an additional regularizing penalty so that  $\nabla_{\mathbf{x},\mathbf{z}}D(\mathbf{x},\mathbf{z})\leq1$. As proved by \citet{donahue2016adversarial} and \citet{dumoulin2016adversarially}, a stable fixed point of this adversarial strategy is for the encoder and decoder to be inverses. 

The discriminator can be interpreted as a learned reward signal for sensory cortex. It signifies in which ways the two distributions meaningfully differ, and tricking the discriminator becomes the purpose of sensory learning. If $p_\theta( \mathbf{x},\mathbf{z})$ and $ q_\phi( \mathbf{x},\mathbf{z})$ are already identical, then $D(\mathbf{x},\mathbf{z})$ cannot find a function that is consistently higher than for $ q_\phi( \mathbf{x},\mathbf{z})$ than $p_\theta( \mathbf{x},\mathbf{z})$ and no learning occurs. 

A major problem with this learning architecture is that the discriminator has an input dimension equal to the total number of neurons. This may prevent stable learning in very large networks. In the ML literature, applications of this algorithm have been limited to dimensions of $\mathbf{x}$ and $\mathbf{z}$ totalling tens of thousands of units at the largest \citep{donahue2019large}. 
In some cases it is possible to decrease the dimensionality by taking into account the structure of the network. This is the case for the application of the method to strictly hierarchical architectures \citep{belghazi2018hierarchical}.

\subsection{Comparing phases of an ascending oscillation}

The cortex arguably switches between modes of processing on multiple timescales. In addition to wake and sleep, oscillations with periods as low as 10ms have also been hypothesized to represent a switch of the main drive of activity from external (bottom-up) to internal (top-down) sources \citep{honey2017switching}. Such oscillations may also be used for adversarial learning (Fig. 2B). Interestingly, this allows for local discriminators with many fewer inputs than in a purely Wake/Sleep algorithm.

An algorithm can be derived by examining the KL divergence $D_{KL}(q\|p)$ for a multilayer network with internal recurrence in each layer. 
In this architecture, we can write the joint log-probabilities as:
\begin{align} \label{recur_DAG}
\log q_\phi( \mathbf{x},\mathbf{z})&=
\sum_i \log q_\phi(\mathbf{z}_i|\mathbf{pa}_i^q)
\hspace{20pt}
\bigg|
\hspace{20pt}
\log p_\theta( \mathbf{x},\mathbf{z})=
\sum_i \log p_\theta(\mathbf{z}_i|\mathbf{pa}_i^p)
\end{align}
These are identical to Equation \ref{eq:DAG-logjointq} except that the single-neuron $z_i$ are replaced with vectors of population activities $\mathbf{z}_i$.
In this case the KL divergence contains a term for each layer $i$, though this term cannot be further factorized into single-neuron terms as in Eq. \ref{eq:KL_MLP}. These terms are the KL divergence between the generative distribution and inference distribution over $\textbf{z}_i$:
\begin{equation}\mathop{\mathbb{E}}_{  q_\phi(\mathbf{x}, \textbf{z})}
\big[\log\frac{q_\phi(\textbf{z}_i|\textbf{z}_{i-1})}{p_\theta(\textbf{z}_i|\textbf{z}_{i+1})}\big]
=\mathop{\mathbb{E}}_{ \textbf{z}_{i-1},\textbf{z}_{i+1}\sim q_\phi(\mathbf{x}, \textbf{z})}
\big[
KL(q_\phi(\textbf{z}_i|\textbf{z}_{i-1})\|p_\theta(\textbf{z}_i|\textbf{z}_{i+1})
\big]
\end{equation}

If these layerwise KL divergences can be minimized, the overall $D_{KL}(q_\phi\|p_\theta)$ will decrease as well.
Instead of using one discriminator that sees the entire network state, then, the brain could use one discriminator per layer that only observes the local population. In the W-GAN formalism, the objective maximized by the discriminator and minimized by the layer would be:
\begin{equation}\label{O_z}
\mathcal{L}_{O, \textbf{z}_i}=
\mathop{\mathbb{E}}_{ q_\phi( \mathbf{x},\mathbf{z})}\bigg[D_i(\mathbf{z}_i)\bigg]
-\mathop{\mathbb{E}}_{q_\phi(\textbf{z}_{i+1}|\mathbf{x})}
\mathop{\mathbb{E}}_{ p_\phi( \textbf{z}_i|\textbf{z}_{i+1})}\bigg[D_i(\mathbf{z}_i)\bigg]
\end{equation}

Evaluating this objective requires obtaining samples of activity in each phase. While the inference distribution is straightforward to sample, the generative distribution $
p_\theta(\textbf{z}_i|\textbf{z}_{i+1}), \hspace{2pt}\textbf{z}_{i+1}\sim q_\phi$ requires an oscillation in which the neurons $\textbf{z}_{i}$ are driven by generative feedback from one layer up. Thus the discriminator would observe alternating phases of somatic activity, with one phase corresponding to bottom-up inputs and one phase corresponding the top-down prediction. 

What we have described here is essentially a stacked version of the VAE-GAN  \citep{larsen2016autoencoding}. This influential work established that the objective $\mathcal{L}_{O, \textbf{z}_i}$ does align the distributions but often results in poor reconstructions. Adding a per-neuron reconstruction error like the mean-squared error, however, implies the problematic assumption that  $p_\theta(\textbf{z}_i|\textbf{z}_{i+1})=\prod_j p_\theta(z_{ij}|\textbf{z}_{i+1})$. As a solution, they propose that the intermediate features of the discriminator could be used as a learned metric for reconstruction distance. This adaptive reconstruction error $\mathcal{L}_{RECON}$ is applied to each layer's reconstruction.

\paragraph{Learning on two timescales:}
Learning can occur on both timescales, and with both algorithms. While one could define a separate discriminator each algorithm, it works well to combine objectives on the same set of discriminators. Defining a hyperparameter $\gamma$ that balances between the objectives, we obtain a combined minimax objective for the $i$th layer and its discriminator:
\begin{equation}\label{eq:gamma}
\mathcal{L}_{\textbf{z}_i} = \gamma \mathcal{L}_{WS} + (1-\gamma)(\mathcal{L}_{O,\textbf{z}_i}+\mathcal{L}_{RECON}).
\end{equation}

\section{Biological characteristics of this adversarial algorithm}

Our broadest hypothesis is that variational Bayesian inference is indeed an objective of sensory learning and that brain areas have meaningful recurrence.
More narrowly we posit that brain areas switch between externally-driven inference and internal self-generation on multiple timescales, that learning aligns probability distributions of activity between phases, and alignment is mediated by a network of teaching signals that adapt to discriminate between phases.

\subsection{Feedback drives neurons in rhythmic cycles}

Recent reviews have argued that particular phases of sleep indeed represent samples from a generative model \citep{hobson2012waking,aru2020apical}, and the idea has long been popular in computational theory \citep{ackley1985learning, hinton1995wake}. It is important to note, however, that this interpretation of sleep remains speculative.

Oscillations, such as alpha in neocortex and theta in hippocampus, have also been argued to represent switches between an external `bottom-up' mode and an internal `top-down' mode \citep{honey2017switching}. Furthermore, \citet{honey2017switching} argue that the computational purpose of such fluctuations is to update internal models of the world to match new observations. In this sense our hypothesis is a specific proposal of how the brain might implement the general principal proposed by \citet{honey2017switching}.

If such oscillations represent mode switches, then one should observe stronger oscillations in activity for unexpected stimuli. After all, for expected (i.e. well-modeled) stimuli the top-down predictions are indistinguishable from bottom-up signals. We thus expect to see the power of the oscillation correlate with surprise, as is observed for the gamma oscillations. For example, in auditory cortex, MEG and EEG studies show increases in gamma power to unexpected auditory stimuli \citep{haenschel2000gamma,todorovic2011prior}, to omissions of expected musical beats \citep{fujioka2009beta}, and to unexpected mismatches between auditory and visual cues \citep{arnal2011transitions}. In the hippocampus, both theta and gamma ranges increase after unexpected stimuli \citep{axmacher2010intracranial}. The correlation of prediction errors with oscillation strength is consistent with the overall framework.

The adversarial interpretation of oscillations makes two further predictions. First, the oscillations should ascend up sensory hierarchies. Indeed, oscillations in the gamma range have been found to ascend up the visual hierarchy \citep{van2014alpha}. Second, top-down feedback in the internal mode should drive somatic activity. Other algorithms for Bayesian inference, such as that in Section \ref{sec:acyclic}, require that feedback only be integrated in dendritic compartments for comparison with bottom-up activity  \citep{siegel2000integrating, friston2005theory}. In reality, feedback into apical dendrites can have a large effect on somatic activity \citep{larkum1999new}.

\subsection{Observable properties of a discriminator}

What would the signals of a discriminator look like? Studies of the internal responses of discriminators indicate they may not be visually distinguishable from other sensory processing neurons \citep{radford2015unsupervised}. It would be difficult to identify one from observations of activity alone. However, in limited circumstances, the optimal discriminator may appear as a surprise signal. If simple set of stimuli have been perfectly learned, the predictive distribution should perfectly match the distribution of evoked activity. If a new stimulus is then shown from a novel distribution of stimuli, that will evoke a pattern of activity unlike any seen before. A good discriminator will increase its activity and appear like surprise. This scenario only occurs when the animal has indeed modeled the stimuli and that the discriminator is optimal.


The connectivity of a discriminator can further narrow candidates.
A purely wake/sleep discriminator receiving input from all neurons would work well in small nervous systems. In the neocortex, then, we hypothesize local discriminators in each cortical column that classify oscillations. Due to their local connectivity these would resemble interneurons with control over local plasticity, especially in the critical period. Somatostatin-positive interneurons meet this rather broad criterion \citep{Yaeger2019}, as do 5-$HT_{3A}R^+$ cells in Layer 1 \citep{takesian2018inhibitory} and likely many others. 

The learning rules implied by our algorithm are perhaps the most distinguishing feature. Connections onto local discriminatory interneurons should have a plasticity rule that switches polarity with the phase of processing. Likewise, the way in which their activity affects plasticity in surrounding cortex should switch polarity depending on the phase of processing.
There is evidence for phasic switches in plasticity in the brain. In the hippocampus, for example, transitions between potentiation and depression and have been observed with phases of the theta rhythm in hippocampus \citep{pavlides1988long, huerta1995bidirectional, hyman2003stimulation}. Synaptic strengths across cortex are also known to homeostatically increase during wake and decrease during sleep \citep{hengen2016neuronal, gonzalez2018activity, pacheco2021sleep}. Future experiments could test whether these or other phasic plasticity rules implement an adversarial algorithm by perturbing patterns of presynaptic neurons in a single phase.

\subsection{Relation to acetylcholine}

The cholinergic system may be a compelling candidate for a discriminator. One of acetylcholine's (ACh) interpretations regards its role in surprise and uncertainty \citep{Yu2005}. Interestingly, like a discriminator, it signals surprise only after some period of learning about typical errors; an `expected uncertainty' \citep{Yu2005}. ACh also has profound control over cortical plasticity during the critical period \citep{Gu2002,Rasmusson2000, Yaeger2019}.
However, the connectivity of cholinergic neurons is not what would be expected of a neocortical network of local discriminators. While the nucleus basalis (the predominant source of cholinergic axons) projects throughout the cortex, it does not receive direct reciprocal connections from sensory cortex \citep{mesulam1984neural}. 
Acetylcholine may instead act to induce switching between internally- and externally-driven phases \citep{hasselmo2006role,honey2017switching,aru2020apical,suzuki2020general}. 
This well-known interpretation is consistent with its its control over sleep stage \citep{OzenIrmak2014} and its control over cortical rhythm \citep{yoder2005involvement,newman2012cholinergic}. For the adversarial hypothesis, such neuromodulatory control is important: discriminator and adversary must agree on the phase. If ACh plays this coordinating role, it should bidirectionally modulate discriminator-dependent learning in the surrounding network as well as onto the discriminator itself.

\subsection{Possible experiments}
While phenomena such as hippocampal replay show that activity in wake and sleep are closely related \citep{nadasdy1999replay}, little is known about the statistical distributions of brain activity during sleep. A study investigating this could closely resemble \citet{Berkes2011a}, which compared the distribution of activity in the ferret visual cortex when eyes are open and closed. Instead of comparing spontaneous closed-eye activity, one could compare activity during phases of LFP oscillations or during sleep. These distributions should align over development. A positive outcome in this experiment would extend indirect evidence that these internally-biased phases produce samples from an internal model \citep{hobson2012waking,honey2017switching}.

Perturbations would allow a deeper test. During the critical period of sensory learning, one could selectively silence activity during a stage of sleep or half of an oscillatory cycle, perhaps via optogenetic methods \citep{andrasfalvy2010two}. This perturbs the generative distribution. One could then observe if activity changes in the waking or opposing phase to match that perturbed distribution. A search for what mediates this alignment could provide a mechanism for adversarial sensory learning.

\begin{figure}[t] 
\includegraphics[width=\linewidth]{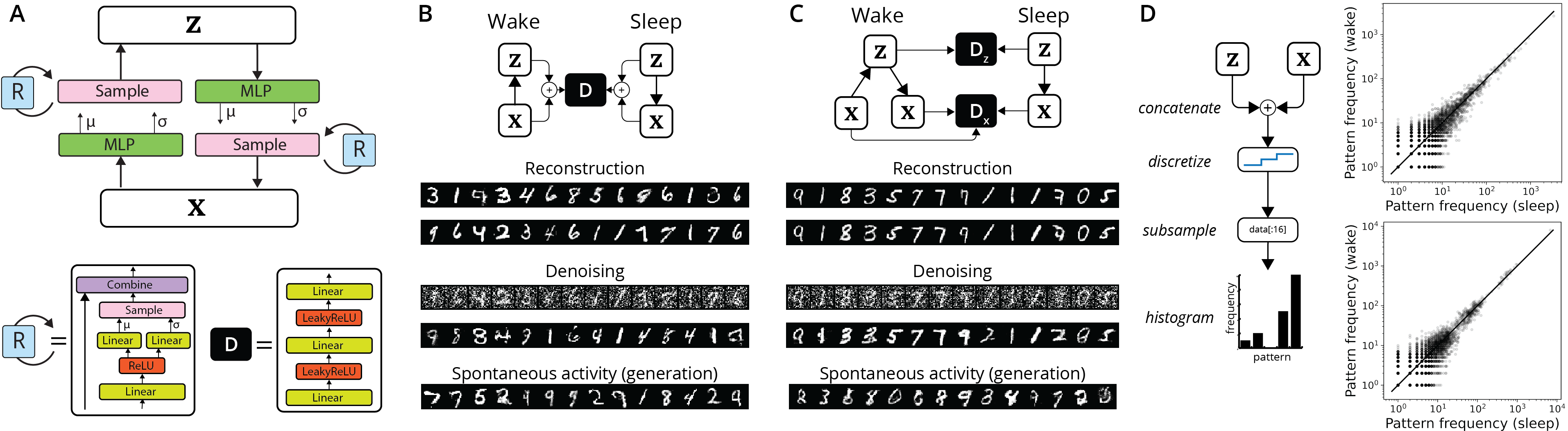} 
\centering
\caption{Generating MNIST digits with a recurrent, stochastic autoencoder.
\textbf{A)} In either inference or generation, samples are passed through a multilayer network parameterizing a Gaussian, which is then sampled and subject to nonlinear stochastic recurrence.
\textbf{B)} In the adversarial wake/sleep algorithm (Section 3.1), a global discriminator observes both $\textbf{z}$ and $\textbf{x}$ during wake and sleep phases. This algorithm allows realistic generation but often shows poor reconstruction.
\textbf{C)} When the discriminator aligns inference with layer-wise reconstructions (Section 3.2), discriminators are local to each layer. We trained this system with the combined objective of Eq. \ref{eq:gamma} with $\gamma=0.1$.
\textbf{D)} As in \citet{Berkes2011a}, we can check empirically if the joint distributions of the Sleep and Wake phases match. After quantizing the vector of $\mathbf{x}$ and $\mathbf{z}$ by rounding to the nearest integer, we observed patterns over $16$ units' activity and histograms of how often these patterns appeared when observing/generating MNIST digits. Comparisons are shown for the model in panel B (top) and in panel C (bottom). 
}
\label{fig:MNIST}
\end{figure}
\section{Simulation Experiments}
To illustrate our proposal, here we apply the learning algorithm presented above. The overall framework of adversarially learned inference over a generative model has been validated elsewhere in the context of feedforward networks \citep{donahue2016adversarial,dumoulin2016adversarially,pu2017adversarial,srivastava2017veegan,larsen2016autoencoding, belghazi2018hierarchical}. As a crucial test of biological feasibility, here we will empirically ask how well the algorithm will work on recurrent architectures.

The Pytorch code used for all figures in this manuscript is available at https://github.com/KordingLab/adversarial-wake-sleep.

\subsection{MNIST digit generation from stochastic recurrent representations}

We begin with a simple autoencoder in which both the encoder and decoder are stochastic, nonlinear transformations followed by recurrence (Fig. \ref{fig:MNIST}). The first-stage transformations are 2-layer fully-connected neural networks that output the mean and variance of a diagonal Gaussian over the latent $\mathbf{z}$ or inputs $\mathbf{x}$, which is then sampled. During recurrence this sample is given to a stochastic network of the same architecture, the output of which is merged with the original sample via an arithmetic mean. As a result of the nonlinearity and interpolation, the conditional distributions $q_\phi(\mathbf{z}|\mathbf{x})$ and $p_\theta(\mathbf{x}|\mathbf{z})$ are not Gaussian. This prohibits evaluating $D_{KL}(q\|p)$ exactly as in a traditional VAE.

We found that recurrence exacerbates a problem with the purely Wake/Sleep approach. While generation is of good quality, reconstructing digits produces different digits than those input (Fig. \ref{fig:MNIST}B). Reconstruction maps to the manifold of MNIST digits, but elsewhere. This is despite the optimal solution being perfect inversion. A wide search over learning rates, weight decay, and the $\beta$ parameters of the Adam optimizer did not produce better reconstructions.

The oscillatory algorithm was much more stable (Fig. \ref{fig:MNIST}C). In addition to the basic adversarial loss ($\mathcal{L}_O$, Eq. \ref{O_z}), we also employed the approach of the VAE-GAN of using the hidden layers of the discriminator as a metric between inputs and reconstructions \citep{larsen2016autoencoding}. This objective resulted in good reconstructions, signifying that the inference network maps to the support of the generative posterior.

To measure the success beyond visual inspection, we employed the approach of \citet{Berkes2011a} and empirically quantified the overlap of the wake and sleep joint distributions. After quantizing the inputs and latents to the nearest integer and subsampling to 16 units (8 $\mathbf{x}$ and 8 $\mathbf{z}$), we counted how many times each unique pattern over the 16 units appeared in wake or sleep. This approximate approach is necessary as the log-likelihood is not tractable due to recurrence. In Fig. \ref{fig:MNIST}D it can be seen that both algorithms are sufficient to align the joint distributions as closely as in \citet{Berkes2011a}.

\begin{SCfigure}[][t]\label{fig:cifar}
\includegraphics[width=.7\linewidth]{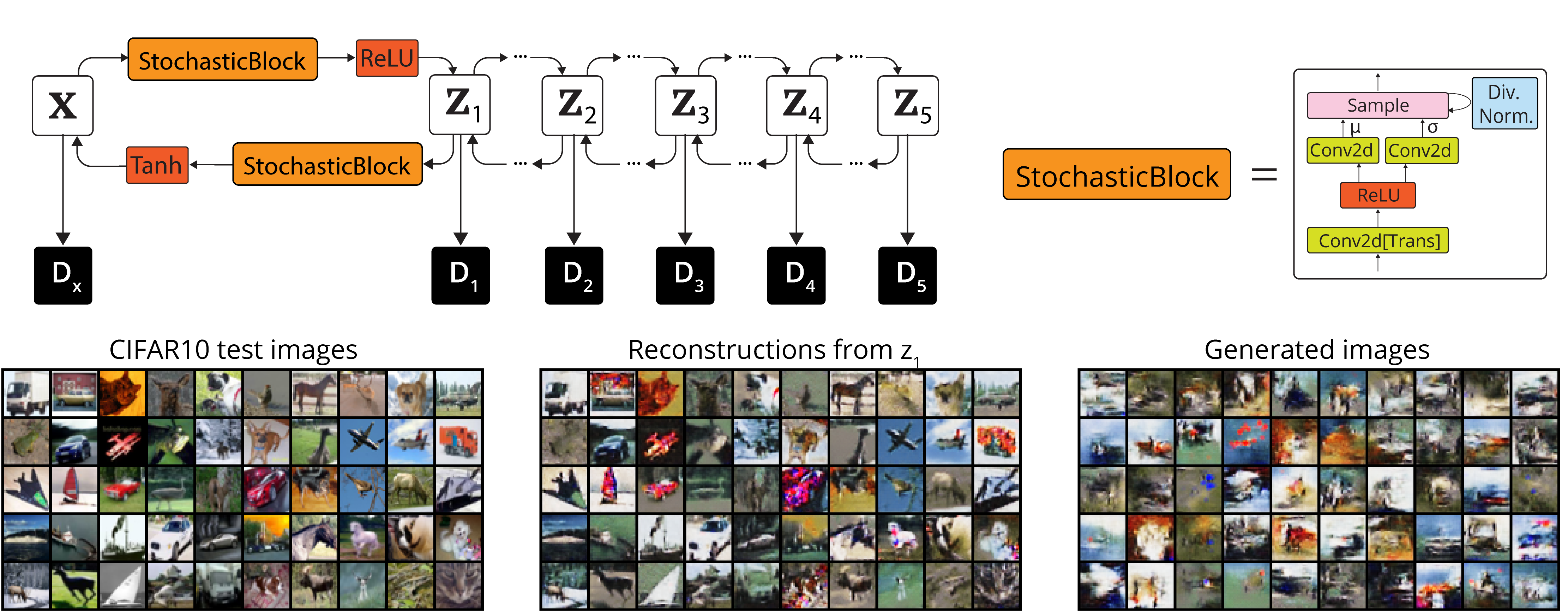} 
\caption{Learning to model images from CIFAR-10 with a hierarchical architecture. Each layer is a stochastic convolutional neural network subject to divisive normalization. We trained towards Eq. \ref{eq:gamma} with $\gamma=0.5$. }
\end{SCfigure}

\subsection{Modeling CIFAR-10 with a hierarchical model}

To test hierarchical models we trained a stochastic variation of the DCGAN architecture using the combined objective of Eq. \ref{eq:gamma} (Fig. \ref{fig:cifar}). Each of the five layers of latent variables and the input is paired with its own discriminator. The conditional distribution of each layer is a reparameterized diagonal Gaussian, but passed through a ReLU nonlinearity and then subject to divisive normalization over each spatial location, a ubiquitous form of local recurrence in the cortex \citep{heeger1992normalization} that is also known to stabilize GAN training \citep{karras2017progressive}. We trained with the combined objective with $\gamma=0.5$; additional model and training details can be found in the Appendix. After 400 epochs, inference again maps to the support of the generative posterior as evidenced by reconstructions (Fig. \ref{fig:cifar}B). Thus, a purely adversarial approach is feasible for deeper networks and more naturalistic datasets.

\begin{figure}[b] 
\includegraphics[width=\linewidth]{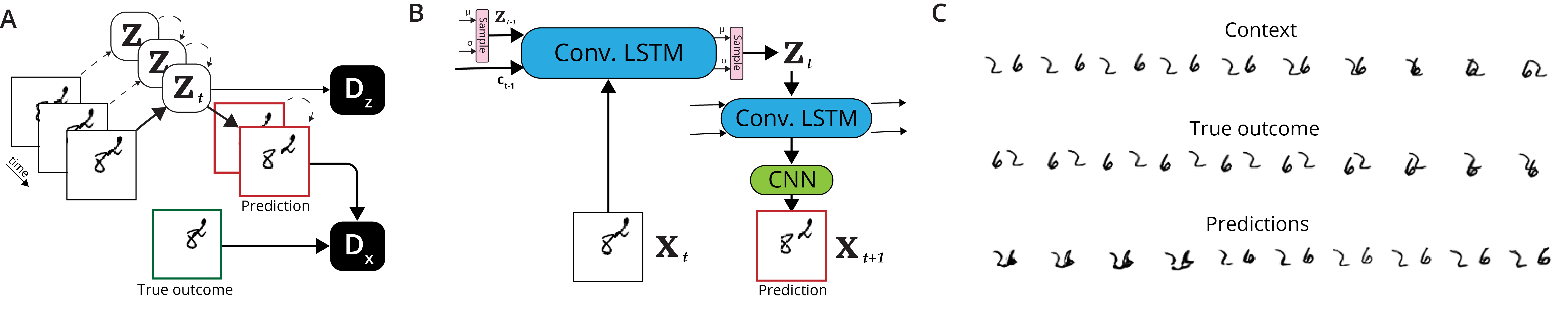} 
\centering
\caption{
\textbf{A)} We trained a recurrent, autoencoding architecture to predict future frames in the Moving MNIST task. Discriminators align inputs and hidden state distributions. \textbf{B)} Encoding and decoding architecture. Hidden states transfer stochastically. \textbf{C)} Conditioned on ten context frames, the trained network generates plausible digits and trajectories.
}
\label{fig:MNIST_vid}
\end{figure}
\subsection{Predicting future inputs}

Recurrence has a clear benefit when inputs vary in time, and as a model of brain processing the temporal dimension is unavoidable. Our framework can readily apply to this setting. Because of conduction delays in real neurons, however, the goal of a loop of processing changes from autoencoding to predicting the future inputs.

Adversarial approaches to video prediction are common in the generative literature. Directly applying the oscillatory algorithm to video prediction, in fact, nearly produces the algorithm proposed in Stochastic Adversarial Video Prediction (SAVP; \citet{lee2018stochastic}). Like the VAE-GAN from which it derives, SAVP regularizes $q_\phi(\mathbf{z}_t)$ to the prior $p(\textbf{z})$ by approximating it as a diagonal Gaussian, and additionally applies an $L_1$ loss over $\mathbf{x}_t$. Both imply the assumption that the conditional distributions $q_\phi(\mathbf{z}_t|\mathbf{x}_{0:t})$ and $p_\theta(\mathbf{x}_{t+1}|\mathbf{z}_{0:t})$ can factorize into single-neuron terms. Here, we demonstrate that the algorithm works when all training is adversarial and the distributions are allowed to be non-Gaussian.

We trained a stochastic architecture (Fig. \ref{fig:MNIST_vid}B) to predict the future frames of video in the Moving MNIST task, in which two digits bounce for 20 frames in a 64x64 area with random initial velocities \citep{srivastava2015unsupervised}. Both the encoding and decoding networks are convolutional LSTMs with stochastic hidden state transitions.
Additional architectural and training details can be found in the Appendix. 

We used separate discriminators for $\mathbf{x}$ and $\mathbf{z}$ to align the joint distribution of latent vectors and subsequent inputs, $(\mathbf{x}_{t},\mathbf{z}_{t-1})$ via Eq. \ref{eq:gamma}. The similarity between the true outcome $\mathbf{x}_t$ and the prediction $\hat{\mathbf{x}}_t$ were also minimized using the hidden layers of the discriminator as a metric. After training, the encoding and decoding are able to predict plausible future video frames (Fig. \ref{fig:MNIST_vid}C).

\section{Discussion}

We have proposed that an adversarial algorithm could provide the missing link between a widely hypothesized computational goal of learning and its implementation in neural systems. The algorithm is compatible with the brain's stochasticity, recurrence, and multilevel architecture, and many of its requirements have already been observed. Several further predictions of the hypothesis are testable with today's technology. 

A central prediction is that the brain meta-learns an objective for sensory learning in the form of a discriminator. This discriminator learns by trying to increase activity during a stage of sleep and decrease activity during wake, or alternatively in the two phases of a fast oscillation. 
This approximate objective is arguably easier to construct for biological areas than directly estimating the variational free energy of the entire sensory cortex, as would be required if meta-learning were not used \citep{rezende2014bio}.

As a model of sensory learning, our hypothesis inherits the limitations of the broader computational goal. The strategy of Bayesian inference over a generative model is under-specified as a complete learning objective because it makes no reference to what subsets of information should be represented by higher sensory areas, or in what form. Learning good representations requires priors over appropriate responses or learned cognitive goals, perhaps embedded in the architecture. The algorithm we propose must act over a supplied architecture and set of priors.

While we have advocated here for variational Bayesian inference over vectors of neural activity, it is worth mentioning that the brain may perform Bayesian inference through other strategies such as particle filtering \citep{lee2003hierarchical} or loopy belief propagation \citep{raju2016inference}. The brain may also infer and generate probability distributions \textit{implied} by the activity of neurons \citep{vertes2018flexible}. 

We assume that the brain contains systems for efficiently minimizing objectives represented in neural activity. In our implementation, all learning depended on the backpropagation of error. Other systems for error minimization may be sufficient for this task. Alternatively, due to a connection between backpropagation and variational autoencoders, there is the possibility that the learned feedforward and feedback connections could themselves could be used for the credit assignment problem \citep{Bengio2014, lillicrap2020backpropagation}.

\subsection{Societal impact}
This work aims to contribute to a greater understanding of the purpose and role of neural systems and cell types. We hope this work will provide a foundation for psychiatric treatment of disorders involving systems for learning and plasticity. Though not our principal aim, this work may also have application to artificial systems which may impact society if further developed. As the constraints on neuromorphic chips are similar to those posed by biology, this work may speed the adoption of those technologies, decreasing the energy demand of AI systems but raising concerns about widespread deployment of AI systems throughout society. If used for lossy video compression, these algorithms could filter information in a way difficult to audit; it is important lossless compression be developed instead. Finally, better unsupervised representation learning may speed the development of intelligent systems and the pace of associated societal change. Like all technology these ideas should be developed benevolently, judiciously, and in the context of responsible governance.

\bibliographystyle{nips_2018}
\bibliography{references}

\section*{Checklist}


\begin{enumerate}

\item For all authors...
\begin{enumerate}
  \item Do the main claims made in the abstract and introduction accurately reflect the paper's contributions and scope?
    \answerYes{}
  \item Did you describe the limitations of your work?
    \answerYes{See Discussion.}
  \item Did you discuss any potential negative societal impacts of your work?
    \answerYes{}
  \item Have you read the ethics review guidelines and ensured that your paper conforms to them?
    \answerYes{}
\end{enumerate}

\item If you are including theoretical results...
\begin{enumerate}
  \item Did you state the full set of assumptions of all theoretical results?
    \answerNA{}
	\item Did you include complete proofs of all theoretical results?
    \answerNA{}
\end{enumerate}

\item If you ran experiments...
\begin{enumerate}
  \item Did you include the code, data, and instructions needed to reproduce the main experimental results (either in the supplemental material or as a URL)?
    \answerYes{}
  \item Did you specify all the training details (e.g., data splits, hyperparameters, how they were chosen)?
    \answerYes{See Appendix}
	\item Did you report error bars (e.g., with respect to the random seed after running experiments multiple times)?
    \answerNA{We included no scalar-valued quantitative figures.}
	\item Did you include the total amount of compute and the type of resources used (e.g., type of GPUs, internal cluster, or cloud provider)?
    \answerYes{See Appendix.}
\end{enumerate}

\item If you are using existing assets (e.g., code, data, models) or curating/releasing new assets...
\begin{enumerate}
  \item If your work uses existing assets, did you cite the creators?
    \answerYes{}
  \item Did you mention the license of the assets?
    \answerYes{See Appendix.}
  \item Did you include any new assets either in the supplemental material or as a URL?
    \answerNA{}
  \item Did you discuss whether and how consent was obtained from people whose data you're using/curating?
    \answerNA{}
  \item Did you discuss whether the data you are using/curating contains personally identifiable information or offensive content?
    \answerYes{None do; See Appendix}
\end{enumerate}

\item If you used crowdsourcing or conducted research with human subjects...
\begin{enumerate}
  \item Did you include the full text of instructions given to participants and screenshots, if applicable?
    \answerNA{}
  \item Did you describe any potential participant risks, with links to Institutional Review Board (IRB) approvals, if applicable?
    \answerNA{}
  \item Did you include the estimated hourly wage paid to participants and the total amount spent on participant compensation?
    \answerNA{}
\end{enumerate}

\end{enumerate}

\newpage
\appendix

\section{Appendix}

\subsection{Training details }

\subsubsection{Software and hardware}

The experiments presented were coded in Pytorch v1.8. All experiments were run on NVIDIA GeForce GTX 1080Ti GPUs. Running on a single card, the experiments take about 5 minutes to train the MNIST architecture, 4 hours to train the CIFAR-10 architecture, and 10 hours to train the Moving MNIST architecture. Searches over training details were performed most for the MNIST and CIFAR-10 tasks and required around 100 times these totals.

\subsubsection{Optimization}

For all tasks we used the Adam optimizer \cite{kingma2014adam} with $\beta_1=0.5$, $\beta_2=0.99$, and weight decay of $2\times 10^{-5}$. The batch size was 512 except for the Moving MNIST task, which for memory reasons was 32. The learning rate of the inference and generation networks was $10^{-4}$ and the discriminator's learning rate of $4\times 10^{-4}$. These learning rates decreased by a factor of $0.96$ between epochs 200 and 250. The total number of training epochs for all tasks was 400. These hyperparameters were chosen from previous literature (specifically \citep{donahue2016adversarial}). Random hyperparameter searches were also conducted, but these did not overly improve results. Architectures were tuned by hand.

For all adversarial objectives we used the Wasserstein-GAN \citep{arjovsky2017wasserstein} with a gradient penalty of $\lambda=1$ \citep{gulrajani2017improved}.

\subsubsection{MNIST architecture}

For the MNIST task \citep{bottou1994comparison} in Figure 3, we used the train set for training and display reconstructions using the test set. MNIST is licensed CC BY-SA $3.0$. The architecture of both the inference and generation networks is a fully-connected, two-layer neural network with 512 hidden units and 128 output units (or 784*2 in the case of generation). The hidden nonlinearities are ReLUs. Half of the output units specify the standard deviation and half the mean of a diagonal Gaussian over the outputs. This is then sampled and fed to nonlinear recurrence with the same architecture as the inference/generation networks. Finally, the output is recombined via an arithmetic mean with the inputs. 

In the global discriminator setup of Figure 3B, the discriminator first concatenated $\textbf{x}$ and $\textbf{z}$, then passed these through a 3-layer fully-connected neural network with linear outputs. Each layer had 512 hidden units and a LeakyReLU unit with negative slope $0.2$. In the local discriminator setup of Figure 3C, each discriminator for $\textbf{x}$ and $\textbf{z}$ was also a 3-layer network with 512 hidden units and LeakyReLU units.

\subsubsection{CIFAR-10 architecture}

For the CIFAR-10 task \citep{krizhevsky2009learning}, available under an MIT licence, we trained an architecture inspired by the DCGAN \citep{radford2015unsupervised}. This dataset does not contain faces or other personal identifying information. The DCGAN is an all-convolutional network with 5 layers. The first layer above the inputs has 32 channels, which doubles every layer, except for the highest layer which has 100 channels. In order for the inference and generator networks to have the same architecture, the generator must use transposed convolutions where the inference uses convolutional operators. This is the standard DCGAN architecture, but where the standard discriminator is used instead as an approximate inference network. 

We modified this network such that in both inference and generation each layer is stochastic. To allow backpropagation through stochasticity, each layer is a reparameterized Gaussian \citep{kingma2013auto}, where the mean and standard deviation are given by a convolutional network with one hidden ReLU layer, linear outputs for the mean, and softplus outputs for the standard deviation. Furthermore, the output of each layer is subject to divisive normalization applied over each pixel dimension (i.e. across channels) as in \citet{karras2017progressive}.

Each of the 5 layers as well as the input are paired with their own discriminator. The 6 discriminators are similar in that they are all neural networks with two convolutional layers followed by a linear output, LeakyReLU activations ($0.2$ negative slope), and pixel-wise divisive normalization on the hidden layers. They differ only in the strides, kernel, and padding of the convolutional layers, which are decreased higher in the network to accommodate the shrinking spatial dimension. For further details about stride and padding, consult the accompanying code.

\subsubsection{Moving MNIST architecture}

Moving MNIST is available under an MIT license \citep{srivastava2015unsupervised}. Our architecture is inspired by \citet{lee2018stochastic} in that encoding and decoding both involve convolutional LSTMs \citep{shi2015convolutional}. We modified the standard deterministic LSTMs such that the output parameterizes a diagonal Gaussian over the hidden state, which is sampled each time the state updates. The stochasticity and nonlinearity implies that over time the distribution of hidden states given the same history of inputs. Our model is one layer deep, and thus has one stochastic hidden state we perform approximate inference over. 

The encoding and decoding LSTMs each have 32 hidden units. The decoding pathway additionally has a two-layer CNN to map the hidden state of the decoding LSTM to the inputs. This CNN has one hidden layer with 16 channels and a ReLU activation.

The discriminator used in this task is a similar architecture as in the CIFAR-10 task. Two convolutional layers with batch normalization and LeakyReLU units are followed by a linear layer, and the features before the linear layer are used as a metric to compare predictions with true outcomes.

Training proceeds as follows. The encoder and decoder LSTMs are first initialized by observing the first 10 frames. During this time period nothing is done with the prediction of the decoder, but hidden states are maintained. Gradients are not cut off to allow backpropagation through time for the future objective. In the second 10 frames, the discriminator over $\mathbf{x}$ attempts to classify predictions $\mathbf{\hat{x}}_t$ from true outcomes $\textbf{x}_t$. The encoder and decoder pathways attempt to trick this, which due to the dependency on previous states requires backpropagation through time. Additionally, the encoder and decoder attempt to reduce the $L_2$ distance of the discriminator's penultimate layer given $\mathbf{\hat{x}}_t$ and $\mathbf{x}_t$.

This solely discriminator-guided prediction method worked well. Interestingly, this places no constraints over $\mathbf{z}$. However, for consistency with the rest of the manuscript, we also trained the discriminator with a sleep phase, enforcing that the distribution over $\mathbf{z}$ matches a prior. The sleep phase involved sampling the hidden state from a prior distribution over $\mathbf{z}$ (a standard normal), and then bouncing back and forth between $\mathbf{x}$ and $\mathbf{z}$ over 10 timesteps using the encoder and decoder pathways. This wake/sleep objective is discounted by $\gamma=0.5$ in the same manner as Eq. \ref{eq:gamma}.

\end{document}